\documentclass[prd,aps,showpacs,preprintnumbers,amssymb]{revtex4}
\usepackage{graphicx}
\usepackage{dcolumn}
\usepackage{bm}
\begin{document}
\title{
Second Order Corrections to QED Coupling at Low Temperature 
}
 \author{Samina S. {\sc Masood}}
 \email{masood@uhcl.edu}
 \affiliation{Department of Physics, University of
Houston Clear Lake, Houston TX 77058}
\author{Mahnaz  {\sc Haseeb}}
 \email{mahnazhaseeb@comsats.edu.pk}
 \affiliation{Physics Department, COMSATS Institute of
Information Technology, Islamabad, Pakistan}
\date{December 2006}

\begin{abstract}
 We calculate the second order corrections to vacuum polarization tensor
of photons at low temperatures, i.e;\ T $\ll 10^{10}$ K ($T << m_e$). The 
thermal 
contributions to the QED coupling constant are evaluated at temperatures below the
electron mass that is $T< m_e$ . Renormalization of QED at these temperatures has
explicitly been checked. The electromagnetic properties of such a thermal medium
are modified. Parameters like electric permittivity and magnetic
permeability of such a medium are no more constant and become functions of temperature.

\end{abstract}
\pacs{11.10 Gh, 11.10.hi, 13.40 Ks, 05.10 Cc, 12.20.-m}
\maketitle
\section{INTRODUCTION}

In quantum field theory, thermal background effects are incorporated
through the radiative corrections. Renormalization of gauge theories
at finite temperature requires the renormalization of gauge
parameters of the corresponding theory. The propagation of particles
and the electromagnetic properties of media are also known to modify
in this framework. One of the ways to obtain these changes is through
the renormalization techniques. Masses of particles are enhanced at
one-loop [1-6], two-loop [7] and presumably to all loop levels.
However, the gauge bosons acquire a dynamically generated mass due
to plasma screening effect [8,9] including the first order radiative corrections in gauge
theories. It helps to determine the changes in electromagnetic
properties [10] of a hot medium. In hot gauge theories $m_{e}$ is the electron
mass and corresponds to $10^{10}$ K. The vacuum polarization tensor in order $\alpha $\ does not
acquire any hot corrections from hot photons in the heat bath [8]
because of the absence of self-interaction of photons in QED.
This effect has already been studied in detail at one-loop level and shown explicitly that electric 
permittivity and magnetic permeability of a medium are modified in real particle background only at higher 
temperatures. However, the higher order corrections to vacuum polarization tensor
of photons are nonzero in the same background. Electric charge therefore
increase in such a situation and consequently leads
to the modifications in electromagnetic properties of a medium due to enhancement in QED coupling.
These statistical contributions are calculated either in Euclidean
or Minkowski space by using imaginary or real-time formalisms,
respectively. In Euclidean space the covariance breaks and time is
included as an imaginary parameter. On the other hand, in real-time
formalism, an analytical continuation of energies along with the
Wick's rotation restores covariance in Minkowski space. It has been
noticed earlier that the breaking of Lorentz invariance can lead to
non-commutative nature of gauge theories [11]. The manifest
covariance is incorporated through the 4-component velocity of
background heat bath which is $u^{\mu }=(1,0,0,0).$

The particle propagators include temperature dependent (hot) term in addition to the
temperature independent (cold) term [1]. At the higher loop level,
the loop integrals involve an overlap of hot and cold terms in
particle propagators. This makes the situation cumbersome and
getting rid of these singularities is a much more involved process.
Sometimes $\delta (0)$ type pinch singularities may appear in
Minkowski space. This problem has been earlier resolved, while calculating [7] the electron self energy
at the two loop level at high temperature where hot
fermion loops contribute. One can get rid of this type of singularities in
thermofield dynamics [12] by doubling the degrees of freedom. However, we do not
come across this type of situation at low temperatures so we do not need to use the
techniques required to handle these singularities due to hot fermion
loops. We work with real-time propagators for our calculations since we are
dealing with the heat bath of real particles for physical systems and keep 
ourselves restricted to on mass shell. The renormalization of QED\ in this 
scheme was checked in detail at one loop level for all ranges of 
temperatures and chemical potentials [5,6,8] of interest. At the higher-loop level, the loop integrals
have a combination of cold and hot terms which appear due to the
overlapping propagator terms in the matrix element. In this paper we calculate the photon
self-energy and restrict ourselves to the low-temperatures, for simplicity, and explicitly
prove the renormalizability of QED at the two-loop level through the order by order cancellation
of singularities.

In the next section we give the detailed calculations of vacuum
polarization tensor at low temperatures. Section III is comprised of
the calculation of second order contributions to QED coupling constant.
The electromagnetic properties of hot media are presented in section IV. 
Discussions of these results are given in section V.

\section{VACUUM POLARIZATION TENSOR IN QED}

Vacuum polarization tensor of photon at the two-loop level gives the
second order hot corrections to charge renormalization constant of
QED at low temperature. This contribution basically comes from the
self mass and vertex type electron loop corrections inside the
vacuum polarization tensor, as in vacuum, the counter term has to be
included to cancel singularities. 

\begin{figure}[!hth]
\begin{center}
  \begin{tabular}{c}
    \mbox{\includegraphics[width=2in,angle=0]
{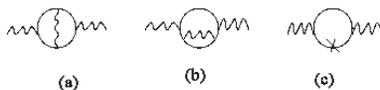}}\\
  \end{tabular}
\end{center}
\caption{Self energy of photon at two loop level} \label{figABC}
\end{figure}

The diagrams in Fig.1 give the
main thermal contributions to the vacuum polarization tensor up to the order $\alpha ^{2}$ at low
temperature, i.e., $T\ll m_{e}$. In this scheme of calculations, the
vacuum polarization tensor of photon in Fig.(1a) is given by

\begin{eqnarray}
\Pi _{\mu \nu }^{a}(p) &=&e^{4}\int \frac{d^{4}k}{(2\pi )^{4}}\int \frac{%
d^{4}l}{(2\pi )^{4}}tr[\gamma _{\mu }S_{\beta }(k)\gamma _{\rho
}D_{\beta
}^{_{\rho _{\sigma }}}(l)S_{\beta }(k+l)  \nonumber \\
&&\gamma _{\nu }S_{\beta }(k+l-p)\gamma _{\rho }S_{\beta }(k-p)],
\end{eqnarray}%
while that in Fig.(1b) is

\begin{eqnarray}
\Pi _{\mu \nu }^{b}(p) &=&e^{4}\int \frac{d^{4}k}{(2\pi )^{4}}\int \frac{%
d^{4}l}{(2\pi )^{4}}tr[\gamma _{\mu }S_{\beta }(k)\gamma _{\rho
}D_{\beta
}^{_{\rho _{\sigma }}}(l)  \nonumber \\
&&S_{\beta }(k+l)\gamma _{\sigma }S_{\beta }(k)\gamma _{\nu
}S_{\beta }(p-k)].\
\end{eqnarray}

The hot terms give the overlapping divergent terms and the removal
of this divergence to establish the renormalization is done by using
specific integration techniques in the rest frame of heat bath.
However, accuracy of the results depend on the order of integration.
This order of integration is so important that the theory can only
be proved renormalizable if the correct order of integration is
maintained. The integration of thermal integrals has to be done
before the temperature independent integrals of cold fermion momenta. We need to get rid of hot 
divergences appearing
due to the presence of the hot boson loops without distributing them
over the cold fermion loops where Lorentz invariance does not break.
Another justification of this order could be found in the fact that
the hot terms in this technique correspond to the contribution of
real background particles on mass-shell and incorporates thermal
equilibrium. However, the preferred frame of heat bath affects the
energy integration of the loop. So the renormalization can only be
proven with the right order of integration. For on-shell
contributions, we do not have to include the off-diagonal elements
of the propagator matrices in our calculations. Moreover, the other
components of the matrices can be related to 1-1 components in
thermofield dynamics.

We restrict ourselves to low temperatures where hot fermion
contribution in background is suppressed and only the hot photons
are contributing from the background heat bath. Therefore, cold
fermion and hot photon propagators are included. The calculations
are simplified if the temperature dependent integrations are
performed before the temperature independent ones. The cold loops
can then be integrated using the standard techniques of Feynman
parametrization and dimensional regularization as is done in vacuum
and are discussed in the standard textbooks [12]. As an illustration
of the importance of the correct order of integration, we just
compare the results for one of these terms. Let us consider the
singular terms when the hot loop in Fig.(1b) is evaluated before the
cold one, we simply get

\begin{equation}
g^{\mu \nu }\Pi _{\mu \nu }^{b}(p,T)=\frac{\alpha ^{2}T^{2}}{3}(1-\frac{2}{%
\varepsilon }),
\end{equation}%
whereas, in the same term, the evaluation of the hot loop after the
cold one gives

\begin{eqnarray}
\bigskip g^{\mu \nu }\Pi _{\mu \nu }^{b}(p,T) &=&-\frac{\alpha ^{2}}{\pi ^{2}%
}[\frac{4\pi ^{2}T^{2}}{3\varepsilon }-\frac{\pi ^{2}T^{2}}{5}  \nonumber \\
&&-\frac{2T^{3}}{5m^{2}}\zeta \left( 3\right) (3|\mathbf{p}|+\frac{49}{3}%
p_{_{0}}+\frac{52p_{0}^{3}T^{4}}{5m^{4}|\mathbf{p}|})].
\end{eqnarray}%
The details of the calculations for Eqs. (3) and (4) are given in
Appendix. The $\frac{1}{\varepsilon }$ contribution is due to the
cold loop momentum integration overlap and is cancelled by the
counter term from Fig. (1c). Calculation of thermal corrections to
self energy of photons becomes much more cumbersome at high
temperatures. However, it has to be done to determine the
corresponding changes in electromagnetic properties of hot media.
First order contributions have already been obtained even for dense
media in this scheme [5,6,10].

Due to the imposed covariance in the propagators, physically
measurable couplings can be evaluated through the contraction of
vacuum polarization tensor $\Pi _{\mu \nu }$ with the metric in
Minkowski \ space $g^{_{\mu \nu }}$ and the bath velocities
$u^{_{^{\mu }}}u^{_{\nu }}$ [8]. This helps to evaluate the
longitudinal and transverse components of vacuum polarization
tensors in order to obtain the thermal contribution to electric
permittivity and magnetic permeability. The gauge invariant finite
contributions of the sum of all the diagrams of Fig.(1) are

\begin{equation}
u^{\mu }u^{\nu }\Pi _{\mu \nu }(p,T)=-\frac{2\alpha ^{2}T^{2}}{3}(1+\frac{%
p_{0}^{2}}{2m^{2}}),
\end{equation}%
and\bigskip

\begin{equation}
g^{\mu \nu }\Pi _{\mu \nu }(p,T)=\frac{\alpha ^{2}T^{2}}{3}.
\end{equation}%
Moreover, the longitudinal and the transverse components of the
vacuum polarization tensor that can be calculated from Eqs. (5) and
(6) are\bigskip

\begin{eqnarray}
\Pi _{L}(p,T) &=&-\frac{p^{2}}{|\mathbf{p}|^{2}}u^{\mu }u^{\nu }\Pi
_{\mu
\nu }(p,T)  \nonumber \\
&=&\frac{2\alpha ^{2}T^{2}p^{2}}{3|\mathbf{p}|^{2}}(1+\frac{p_{0}^{2}}{2m^{2}%
}),
\end{eqnarray}%
and\bigskip

\begin{eqnarray}
\Pi _{T}(p,T) &=&-\frac{1}{2}[\Pi _{L}(p,T)-g^{\mu \nu }\Pi _{\mu
\nu
}^{b}(p,T)]  \nonumber \\
&=&\frac{\alpha ^{2}T^{2}}{3}[\frac{1}{2}-\frac{p^{2}}{|\mathbf{p}|^{2}}(1+%
\frac{p_{0}^{2}}{2m^{2}})].
\end{eqnarray}%
respectively. These components of the vacuum polarization tensor are
used to determine the electromagnetic properties of a medium with
hot photons and will be discussed in the next section in a little
more detail. It is worth-mentioning that Fig.(1b) gives major
contributions to the finite terms in the above equations. However,
the other two diagrams are needed to establish the cancelation of
singularities even at low temperature. It is also interesting to
note that the order of singularities is the same as the order of
$\alpha $ in the matrix element. It is not an accidental occurrence
of the matching order of singularity. It is due to the fact that the
number of loops and the number of maximum delta functions associated
with photon propagators in a term in the matrix element are the same
and would correspond to the highest order of hot divergences.

\section{CHARGE RENORMALIZATION}

Vacuum polarization in a medium gives modification to electric
charge and the coupling constant in QED. The electric charge couples
with the medium through vacuum polarization and picks up thermal
corrections accordingly. This change leads to the enhancement in
coupling constant. Using Eq.(8) from the last section and the
standard method of evaluation of charge renormalization constant of
QED [12], the electron charge renormalization up to the order
$\alpha ^{2}$ can be expressed as

\begin{equation}
Z_{3}=1+\frac{\alpha ^{2}T}{6m^{2}}^{2}.
\end{equation}%
The first order temperature dependent term in the above equation is
not present indicating the absence of corrections from one loop as
calculated in detail in Ref. [8]. The corresponding value of the
QED\ coupling constant comes out to be

\begin{equation}
\alpha _{R}=\alpha (T=0)(1+\frac{\alpha ^{2}T^{2}}{6m^{2}}).
\end{equation}%
It can be clearly seen in the above equations that the electron
charge and hence the QED coupling constant goes smaller with the
increased order of loops which clearly assures the renormalization
of electron charge. This change in the coupling constant leads to
changes in the electromagnetic properties of a medium which we will
discuss in the next section.

\section{ELECTROMAGNETIC PROPERTIES OF A MEDIUM}

This is a well-known fact that the electromagnetic properties of
media depend [8] on the coupling of charge with the medium, that is
determined through the statistical properties. One-loop
corrections do not change it at low temperatures whereas the second
order corrections contribute to them. It can be seen that the
electric permittivity of a medium at the two loop level modifies to

\begin{eqnarray}
{\varepsilon }(p,T) &=&1-|\mathbf{p}|^{2}\Pi _{L}(p,T)  \nonumber \\
&=&1-\frac{2\alpha ^{2}T^{2}p^{2}}{3}(1+\frac{p_{0}^{2}}{2m^{2}}),
\end{eqnarray}%
whereas the magnetic permeability is given by

\begin{eqnarray}
\frac{1}{\mu (p,T)} &=&1+\frac{1}{p^{2}}[\Pi _{T}(p,T)-\frac{p_{0}^{2}}{|%
\mathbf{p}|^{2}}\Pi _{L}(p,T)]  \nonumber \\
&=&1+\frac{\alpha ^{2}T^{2}}{3}[\frac{1}{2p^{2}}-\frac{1}{|\mathbf{p}|^{2}}%
(1+\frac{p_{0}^{2}}{2m^{2}})(1+\frac{2p_{0}^{2}}{|\mathbf{p}|^{2}})].\
\end{eqnarray}%
Equations (11) and (12) show that the second order radiative
emission and absorption of hot photons from the heat bath leads to
deviations from unity in the values of the dielectric constant and
the magnetic susceptibility. This particular feature is not present
at the one-loop level at low temperature because of the absence of
thermal corrections to $\Pi _{L}$\ and
$\Pi _{T}$\ for cold fermions. There are two ways to approach \ limit. If $%
p_{0}=|\mathbf{p}|$ in the rest frame of the heat bath and then the limit $|%
\mathbf{p}|\longrightarrow 0$\ is taken, we get

\begin{equation}
\kappa _{L}^{2}\longrightarrow \lim_{|\mathbf{p}|\longrightarrow
0}\Pi _{L}(0,|\mathbf{p}|,T)=\frac{2\alpha ^{2}T^{2}}{3},
\end{equation}

\begin{equation}
\kappa _{T}^{2}\longrightarrow \lim_{|\mathbf{p}|\longrightarrow
0}\Pi _{T}(0,|\mathbf{p}|,T)=\frac{\alpha ^{2}T^{2}}{2}.
\end{equation}%
On the other hand if we set $p_{0}=0$ with
$|\mathbf{p}|\longrightarrow 0$ then we obtain

\begin{equation}
\omega _{L}^{2}\longrightarrow \lim_{|\mathbf{p}|\longrightarrow 0}\Pi _{L}(|%
\mathbf{p}|,|\mathbf{p}|,T)=0,
\end{equation}

\begin{equation}
\omega _{T}^{2}\longrightarrow \lim_{|\mathbf{p}|\longrightarrow 0}\Pi _{T}(|%
\mathbf{p}|,|\mathbf{p}|,T)=\frac{\alpha ^{2}T^{2}}{6}.
\end{equation}%
The difference of the longitudinal and transverse components of
the vacuum polarization tensor\ in $p^{2}\longrightarrow 0$ limit
measures the dynamically generated mass of the photon.
Electromagnetic properties of a medium change due to this
perturbative mass.

\section{RESULTS\ AND DISCUSSIONS}

We have previously noticed [8] that the low temperature effects on
vacuum polarization tensor due to hot photon background at one
loop level are zero because of the absence of self couplings of
photons. However, we found in this paper that thermal corrections to
vacuum polarization tensor are non-zero at the higher loop level,
even at low temperature. It occurs due to the overlap of hot photon
loop with the cold fermions loops. It is effectively the mutual interaction
of propagating hot photons through the cold fermions of the medium.
This leads to the modifications in electromagnetic properties of a 
medium itself and are expected to give larger contribution at higher
loop levels. However, most of these terms are finite and the order by order
cancellation of singularities can be observed through the addition
of all the same order diagrams. 
The renormalization of the theory can only be proved if covariant hot integrals are
evaluated before the cold divergent integrals on mass shell. Once
the hot loop energies are integrated out, the usual vacuum
techniques of Feynman parameterization and dimensional
regularization can be applied to get rid of vacuum singularities. It
is also worth-mentioning that all the hot corrections give a similar $T^{2}$
dependence. The incorrect order of integration gives increasing order of $T$ (See Eq.(4)) due to the 
overlap with the vacuum divergences.
This unusual behaviour of hot integrals appear due to the overlap of hot and cold terms.
Whereas, the usual regularization techniques of vacuum theory like dimensional regularization would only 
be
valid in a covariant framework of a lorentz invariant system. Higher order terms may even give a 
stronger dependance on T with this inverted order of integration.

At the two loop level, it has been checked in detail by integration
that the vertex type corrections to the virtual electrons vanish and
the self energy type corrections to the electron loops contribute to
QED\ coupling constant at low temperatures. It indicates the mutual interaction 
of photons at the higher loop level. The dominant temperature corrections in
this hot medium are of the order $\alpha $ \ which implies the
convergence of the perturbative series. Since the thermal effects
are incorporated as perturbative effects, at sufficiently low
energies ($E<<m_{e}$), the finite temperature corrections become
negligible and one can recover the terms in vacuum by taking $T=0$
in these results. The presence of the statistical contribution of
photon propagator modifies the vacuum polarization and hence the
electron charge which leads to changes (though small) in the
electromagnetic properties of the hot medium even at low
temperatures. Mass, wavefunction, and charge of electron are
renormalized in the presence of heat bath. These renormalized
values give the dynamically generated mass of photon and its
effective charge in such a background. Equations (11) and (12) give
the estimation of dielectric constant and magnetic permeability of
a hot medium. Both of these quantities deviate from unity whereas
their value was unity in QED at the one loop level [13].

It may be noted that as $p^{2}\longrightarrow 0,$ Eq. (8) reduces to
\begin{equation}
\Pi _{T}(p)=\frac{\alpha ^{2}T^{2}}{6},
\end{equation}%
which corresponds to the dynamically generated thermal mass of
photon in the hot background. This type of effect has been earlier
observed for self-mass of gluon even at the one-loop level [9] due to the self-coupling of gluons.
With rising temperature this photon mass drastically change the behavior of this coupling, especially when 
T$\geq$ m. Further we obtain the respective propagation vectors and frequencies
when we take the limiting values for the longitudinal and transverse
component of the vacuum polarization tensor in Eqs. (7) and (8). In
particular, in Eq. (13), $\frac{2\alpha ^{2}T^{2}}{3}$  represents
the Debye screening length in such a medium. This helps to evaluate
the decay rates and the scattering crossections of particles in such
media.

It is worth-mentioning that in the real-time formalism, the propagator 
has two additive terms, the vacuum term and the temperature dependent hot 
term. Therefore, in the second order perturbation theory we get purely hot 
term ($~\alpha^2 T^4/m^4$), purely cold terms (~$T^0$) and the 
overlapping hot and cold terms ($~\alpha^2 T^2/m^2$). These terms can only 
be obtained in this formalism at the two-loop level. So the overall result 
comes out to be a 
combination of all of these terms. For example, eq.(10) can be written as 

\begin{equation}
\alpha _{R}=\alpha (T=0)(1+ \frac{\alpha 
^{2}T^{2}}{6m^{2}}+O(\frac{T^{4}}{m^{4}})).
\end{equation}%

Since the last term ($~\alpha^2 T^4/m^4$), is much smaller than the second 
term ($~\alpha^2 T^2/m^2$), so we can ignore 
this term at the moment. We got similar results in
Ref. [7] regarding the self-mass of electron in QED at two loops level.  
However, for $T>m$, the last term has to be evaluated. It was correctly 
speculated in Ref. [16] that the overall effect is of the order $T^4/m^4$. 
In some of the earlier results, the effective potential approach has been 
used to obtain two loop thermal corrections as $T^4/m^4$ [14,15]. 
  Here in this paper, the dominent contribution of the two hot loops is calculated for temperature 
sufficiently smaller than the electron mass. The leading order contribution at low T is based on the perturbative 
expansion in QED and the first term should be proportional to $T^2/m^2$ in this expansion. The damping 
factor $exp(-m/T)$ appears in the effective action when all the contributing terms are included 
simultaneously. We look at all these terms one by one and work for  
sufficiently small values of temperature to get the simple and approximate results to evaluate some 
physically measurable parameters. We also ignore the magnetic field effect in this calculation. 
It is again a reasonable approximation to demonstrate the renormalizability of QED at 
low temperatures. Next term in this calculation is proportional to 
$T^4/m^4$ with a smaller coefficient. However, this term is non-ignoreable and 
is needed to be included at high temperatures. We are already working on the 
evaluation of such terms which have some extra high T singularities. The perturbative analysis in QED is important to show the order by order cancellation of 
singularities in real-time formalism. In this formalism, the leading order terms can be evaluated 
ignoring the rest of it. 
We are now working on the evaluation of the next term. Even though it is a lengthy procedure, a 
good estimate of the background contribution can be obtained from it at 
high T also. 
In QCD, the effective potential approach has to be used since perturbative 
analysis becomes unmanageable due to the self-coupling of gluons in the presence of hot and dense media. 
Such estimations are done in literature [17]. However, this effective potential approach has to be used in QED
 and other theories, especially for the calculation of the effective action.

\section{APPENDIX}

The hot contribution at low temperature, comes from the photon
background only. Eq. (2) for Fig. (1b) therefore simplifies to

\begin{eqnarray}
\Pi _{\mu \nu }^{b}{(p)}{=\Pi }_{\mu \nu }^{b}{(p,T=0)} &&-{2\pi
ie}^{4}\int
\frac{d^{4}k}{(2\pi )^{4}}\int \frac{d^{4}l}{(2\pi )^{4}}N_{\mu \nu }^{b}{%
\delta (l}^{2}{)n}_{B}{(l)}  \nonumber \\
\times &&\frac{1}{(k^{2}-m^{2}+i\varepsilon )^{2}\left[ (p-k)^{2}-m^{2}+i%
\varepsilon \right] \left[ (k+l)^{2}-m^{2}+i\varepsilon \right] },\
\ \
\end{eqnarray}

where%
\begin{eqnarray}
N_{\mu \nu }^{b} &=&8\left[ \left( 3m^{2}-k^{2}-2k.l\right) \right]
k_{\mu
}(p-k)_{\nu }  \nonumber \\
&&+k_{\nu }(p-k)_{\mu }-g_{\mu \nu }k.(p-k)+\left(
k^{2}-m^{2}\right)
\{l_{\mu }(p-k)_{\nu }  \nonumber \\
&&+l_{\nu }(p-k)_{\mu }-g_{\mu \nu }l.(p-k)\}+2g_{\mu \nu
}m^{2}(m^{2}-k.l)],\ \
\end{eqnarray}%
\ \ \ \ \ \ \ \ \ \ \ \ \ \ \ \ \ \ \ \ \ \ \ \ \ \ \ \ \ \ \ \ \
\bigskip\ \ \ \ \ \ \ \ \ \bigskip\ \ \ \ \ \ \ \ \ \ \
\begin{equation}
g^{\mu \nu }\Pi _{\mu \nu }^{b}(p,T)=\frac{16e^{4}}{(2\pi )^{6}}\int \frac{%
\left( 2p.k-k^{2}-m^{2}\right) d^{4}k}{\left( k^{2}-m^{2}\right)
^{2}\left( k^{2}-m^{2}-2p.k\right) }\int
|\mathbf{l}|d|\mathbf{l}|n_{B}(l),
\end{equation}%
with

\begin{equation}
\int |\mathbf{l}|d|\mathbf{l}|n_{B}(l)=\frac{\alpha ^{2}T^{2}}{6}.
\end{equation}%
\ \ Now integrating over the cold loop by using Feynman
parametrization and dimensional regularization one gets Eq. (3). On
the other hand, first doing Feynman parametrization, integrating by
dimensional regularization, and simplifying one gets \ \
\begin{eqnarray}
g^{\mu \nu }\Pi _{\mu \nu }^{b}(p,T) &=&  \nonumber \\
&&\frac{e^{4}}{8\pi ^{5}}\int d^{4}l\delta
(l^{2})n_{B}(l)\int_{0}^{1}dx\smallskip \int_{0}^{x}dy\int_{0}^{y}dz
\nonumber \\
&&[\frac{1}{\{-m^{2}-2p.ly(y+z)\}^{2}}%
\{-2(p.l)^{2}(2y^{4}+2y^{2}z^{2}+4y^{3}z-y^{2}-yz)\smallskip  \nonumber \\
&&+m^{2}(p.l)(6y^{2}+6yz-y+z)+4m^{4}\}  \nonumber \\
&&+\frac{\left( p.l\right)
(24y^{2}+24yz-18y-12z)-6m^{2}}{m^{2}-2p.ly(y+z)}
\nonumber \\
&&-12\{\frac{1}{\eta }-\gamma -\ln (\frac{4\pi
}{-m^{2}-2p.ly(y+z)})\}],
\end{eqnarray}%
\ \ and%
\begin{eqnarray}
g^{\mu \nu }\Pi _{\mu \nu }^{b}(p,T) &=&  \nonumber \\
&&\frac{e^{4}}{8\pi ^{4}}\int d\mu \int |\mathbf{l}|d|\mathbf{l}%
|n_{B}(l)\int_{0}^{1}dx\smallskip \int_{0}^{x}dy\int_{0}^{y}dz  \nonumber \\
&&\{2(2y^{4}+2y^{2}z^{2}+4y^{3}z-y^{2}-yz)|\mathbf{l}|^{2}\smallskip
\lbrack \frac{\left( p_{_{0}}-|\mathbf{p}|\mu \right)
^{2}}{a_{+}^{2}}+\frac{\left(
p_{_{0}}+|\mathbf{p}|\mu \right) ^{2}}{a_{-}^{2}}]  \nonumber \\
&&+m^{2}|\mathbf{l}|(6y^{2}+6yz-y+z)[\frac{\left(
p_{_{0}}-|\mathbf{p}|\mu
\right) }{a_{+}^{2}}-\frac{\left( p_{_{0}}+|\mathbf{p}|\mu \right) }{%
a_{-}^{2}}]  \nonumber \\
&&+|\mathbf{l}|(24y^{2}+24yz-18y-12z)[\frac{\left(
p_{_{0}}-|\mathbf{p}|\mu
\right) }{a_{+}^{2}}-]  \nonumber \\
&&+4m^{4}[\frac{1}{a_{+}^{2}}+\frac{1}{a_{-}^{2}}]-6m^{2}\left[ \frac{1}{%
a_{+}}+\frac{1}{a_{-}}\right]  \nonumber \\
&&-12\left[ 2(\frac{1}{\eta }-\gamma -\ln 4\pi )-\ln \left(
a_{+}\right) -\ln \left( a_{-}\right) \right] \},
\end{eqnarray}%
where \ \ \ \ \ \ \ \ \ \ \ \ \ \ \ \ \ \ \ \ \ \ \ \ \ \ \ \ \ \ \
\ \ \ \ \ \ \ \ \ \ \ \ \ \ \ \ \ \ \ \ \ \ \ \ \ \

\begin{equation}
a_{\pm }=m^{2}\pm 2y(y+z)\left( p_{_{0}{}^{-}+}|\mathbf{p}|\mu
\right) .
\end{equation}%
After a bit lengthy calculation of the integrals, these relations
finally give Eq. (4).

\section{REFERENCES}

\begin{enumerate}
\item J. F. Donoghue and B. R. Holstein, Phys. Rev. \textbf{D28}, 340
(1983); J. F. Donoghue and B. R. Holstein, Phys. Rev. \textbf{D29}, 3004E (1983); J.F.
Donoghue, B. R. Holstein, and R. W. Robinett, Ann. Phys. (N.Y.)
\textbf{164}, 233 (1985).

\item A.Weldon, Phys. Rev. \textbf{D26}, 1394(1982); A.Weldon, Phys. Rev. \textbf{D26}, 2789
(1982); C. Bernard, Phys. Rev. \textbf{D9}, 3312 (1974); W.Dittrich, Phys.Rev. \textbf{D19}, 2385 (1979); P.H. Cox, W.S. Hellman and A.Yildiz, Ann. 
Phys. 154, 211 (1984). 

\item L. Dolan and R. Jackiw, Phys. Rev. \textbf{D9}, 3320 (1974); A.J.Niemi
and Semenoff, Nucl. Phys. \textbf{B230} [FS10], 181 (1984).

\item P. Landsman and Ch G. Weert, Phys. Rep. \textbf{145}, 141
(1987) and the references therein. 

\item K. Ahmed and Samina Saleem (Masood), Phys. Rev. \textbf{D35}, 1861
(1987); K. Ahmed and Samina Saleem (Masood), Phys. Rev. \textbf{D35}, 4020(1987).

\item Samina S. Masood, Phys. Rev. \textbf{D44}, 3943(1991); Samina S.
Masood, Phys. Rev. \textbf{D47}, 648(1993); Samina Saleem
(Masood),\ Phys. Rev. \textbf{D36}, 2602(1987).

\item Mahnaz Qader (Haseeb), Samina S. Masood, and K. Ahmed, Phys. Rev.
\textbf{D44}, 3322(1991); Samina S. Masood, Phys. Rev. \textbf{D46}
, 5633(1992).

\item K. Ahmed and Samina S. Masood, Ann. Phys. (N.Y.) \textbf{207},460(1991)
.

\item Samina Masood and Mahnaz Qader (Haseeb), Astroparticle Phys. \textbf{3} 405
(1995).

\item See for example Ref.[8,9], Samina S. Masood, Phys. Rev. \textbf{D48,} 3250(1993); Samina S.
Masood, Astroparticle Physics \textbf{4}, 189(1995); Samina S.
Masood, Proceedings of the Europhysics Conference in Brussels, 524
(1995), and Samina S.\ Masood, \textbf{hep-ph/0109042}.

\item See for example, C. Brouder, A. Frabetti \textbf{hep-ph/0011161} and
F. T. Brandt, Ashok Das, J. Frenkel, Phys. Rev. \textbf{D65}, 085017
(2002).

\item See for example: C. Itzykson and J. B. Zuber, \textit{Quantum Field
Theory }(McGraw- Hill Inc., 1990).

\item Samina S. Masood and Mahnaz Qader (Haseeb), Phys. Rev. \textbf{D46},
5110 (1992); Samina S. Masood and Mahnaz Qader (Haseeb), 'Finite
temperature and density corrections to electroweak decays',
Proceedings of 4th Regional Conference on Mathematical Physics,
Sharif Univ. of Technology, Tehran, IRAN (12-17 May 1990), Eds: F.
Ardalan, H. Arafae, and S. Rouhani, 334, Sharif University of
Technology Press; Samina S. Masood, \textbf{hep-ph/0108126} and
references therein.
\item  J.O. Andersen, E. Petitgirard, M. Stricklanda, 
Phys. Rev. \textbf{D70}, 045001(2004), \textbf{hep-ph/0302069}.

\item  H. Gies Phys. Rev. \textbf{D61}, 085021(2000), 
\textbf{hep-ph/9909500}.
 
\item Duane A.Dicus, David Dawn and Edward W.Kolb, Nucl. Phys.\textbf{B223}, 
525(1981).

\item See for example, F. T. Brandt, \textit{et}.\textit{al.}, \textbf{
hep-th/0601227} and J. Alexandre, Phys. Rev. \textbf{D63}, 073010
(2001).
\end{enumerate}
\end{document}